\documentclass[12pt]{iopart}
\usepackage{amssymb}
\usepackage{iopams}
\usepackage{bm}
\usepackage{epsf,graphicx,color}

\newcommand{\eps}{\varepsilon}

\newcommand{\eff}{{\rm eff}}

\newcommand{\la}{\langle}
\newcommand{\ra}{\rangle}
\newcommand{\lla}{\left\langle}
\newcommand{\rra}{\right\rangle}

\newcommand{\One}{\bm{1}}

\newcommand{\alphaout}{\alpha^{\rm out}}
\newcommand{\alphain}{\alpha^{\rm in}}
\newcommand{\sigmaind}{\sigma_{\rm ind}}
\newcommand{\sigmadir}{\sigma_{\rm dir}}

\begin{document}

\title[Statistical quantum decay]
   {Random matrix description of decaying quantum systems}

\author{T Gorin$^1$}

\address{\hspace*{-1.5mm}\mbox{}$^1$Max-Planck-Institut f\" ur Physik komplexer
   Systeme, N\" othnitzer Str. 38, D-01187 Dresden, Germany}

\begin{abstract}
This contribution describes a statistical model for decaying quantum systems
({\it e.g.} photo-dissociation or -ionization). It takes the interference 
between direct and indirect decay processes explicitely into account.
The resulting expressions for the partial decay amplitudes and
the corresponding cross sections may be considered a many-channel 
many-resonance generalization of Fano's original work on resonance lineshapes 
[Phys. Rev {\bf 124}, 1866 (1961)]. 

A statistical (random matrix) model is then introduced. It allows to describe 
chaotic scattering systems with tunable couplings to the decay channels. We 
focus on the autocorrelation function of the total (photo) cross section, and
we find that it depends on the same combination of parameters, as the 
Fano-parameter distribution. These combinations are statistical variants of 
the one-channel Fano parameter. It is thus possible to study Fano interference 
({\it i.e.} the interference between direct and indirect decay paths) on the 
basis of the autocorrelation function, and thereby in the regime of overlapping 
resonances. It allows us, to study the Fano interference in the limit of 
strongly overlapping resonances, where we find a persisting effect on the 
level of the weak localization correction.
\end{abstract}

\pacs{05.45.Mt, 03.65.Nk}


\ead{gorin@pks.mpg.de}


\maketitle


\section{\label{I} Introduction}

Molecular photodissociation~\cite{Rei94,Dob95,Dob95b,Pes95,Kel96,Kir00}
and atomic autoionization~\cite{GreDel97} are examples of quantum-mechanical
decay processes: Within the dipole approximation, the absorption
of a photon excites the quantum system into an energy region, which allows the
dissociation and/or ionization of the system. Technically speaking, one 
observes the decay of an initial wave packet $|\alpha\ra$ within a scattering 
system, via accessible open channels. Given a scattering system with 
Hamiltonian $H$, the partial decay amplitudes $t_a(E)$ are of central 
importance, as they define the partial and total (photo) cross sections:
\begin{equation}
t_a(E) = \la\Psi^{(a)}(E)|\alpha\ra \qquad \sigma_a(E)= \sigma_0\; |t_a(E)|^2
   \qquad \sigma(E)= \sum_c \sigma_c(E) \; .
\label{I:damp}\end{equation}
Here, $\Psi^{(a)}(E)$ is the {\em post-controlled} scattering solution of the
problem. This means that, asymptotically, there is one single outgoing plane 
wave in channel $a$. The sum in the definition of the total cross section 
$\sigma(E)$ runs over all open channels. The proportionality constant 
$\sigma_0$ depends on the excitation mechanism. As we will not discuss those 
mechanisms, we set $\sigma_0 = 1$. 
This description reduces the whole process to a
{\em half collision processes}~\cite{Rost98,HalCol91,Schinke93}. Note that
we will mostly use the terminology from photo absorption, for simplicity,
being well aware of the fact that other excitation mechanisms are equally 
possible.

We aim at a description of chaotic scattering systems, which have a large
interaction region (in terms of elementary Planck cells), 
but a finite number of open channels. Such a situation occurs frequently
in the photodissociation of simple molecules, where the potential surface
allows trajectories coming from infinity to enter the interaction region
only through so called ``bottlenecks'' (transition states). In this
situation, scattering trajectories may have very short dwell times: 
They either do not find the bottleneck to enter the interaction region, or, 
due to symmetry, leave the interaction region after only a few bounces 
(Ehrenfest time). These are direct processes.
The remaining trajectories ``randomize'' and leave the
interaction region only after the average dwell time, which is proportional
to the size of the openings. They contribute to the indirect decay.

In~\cite{AlhFyo98,FyoAlh98}, Alhassid and Fyodorov 
formulate and solve a statistical model for the total cross section.
This quantity has the convenient property that it
can be expressed in terms of the Greens function~\cite{Hel78a,Hel78b}. With
the help of Feshbach's projection formalism~\cite{Fes58,Fes62}, the full 
Hilbert space is divided 
into an orthogonal sum of a number of quasi bound states, and an $M$-channel
continuum. In~\cite{FyoAlh98}, an analytical expression for the 
autocorrelation function of $\sigma(E)$ is derived, by using the supersymmetry 
method. The result is quite similar to the 
``Verbaarschot-Weidenm\" uller-Zirnbauer'' (VWZ) integral~\cite{VWZ85}. 

In particular dynamical systems, deviations from the statistical model are
quite possible. In the semiclassical regime, for instance, Policot-Ruelle
resonances may give rise to characteristic signatures in the autocorrelation 
function of the total photo cross section~\cite{Aga99,Aga00}. A different 
source for deviations, is the interference with direct decay processes (Fano 
interference). In the limit of isolated resonances, it leads to asymmetric 
lineshapes~\cite{Beu35,Fano61}. The statistical model of Alhassid and Fyodorov
has been generalized recently, to include direct 
processes~\cite{AFGIM03,GMI04}. The basic formalism has been developed in 
nuclear reaction theory ~\cite{Feshbach92} (and references therein); see also
reference~\cite{SRSM97b}. 

Fano interference (Beutler-Fano lineshapes) has also been studied in mesoscopic
transport. It is shown in reference~\cite{CWB01} that Fano interference can be 
observed in the conductance of quantum dots coupled to single mode leads. 
Even though these results are not simply related to the present model, they 
still point at important applications. Quite generally, a Fano resonance may 
be used as a kind of built-in interferometer, which, for instance, is very 
sensitive to decoherence effects~\cite{Kob04}. Also temperature effects as 
well as Coulomb interaction effects can be studied~\cite{Joh04}.\\

In this paper, we address the question how to detect and quantify Fano 
interference in the many channel case, and in the regime of overlapping 
resonances. For that purpose, we review in section~\ref{F} the statistical 
model for half collision processes, as developed in~\cite{AFGIM03,GMI04}. 
The mapping formalism of reference~\cite{GMI04}, together with an important
generalization is presented in section~\ref{M}. We then consider the limit of 
isolated resonances, where the resonance lineshapes can be described by a 
complex Fano parameter (section~\ref{L}). The statistics of the Fano parameter 
has been considered in~\cite{Ihra02}, for the one channel case. 
Here, we identify those quantities, which determine the Fano-parameter 
distribution in the many channel case. For each
channel, one can find a statistical analog of the one-channel Fano parameter.
In section~\ref{A}, we turn to the study of the autocorrelation function. 
Surprisingly, it depends on very similar quantities. This allows to quantify 
Fano interference, beyond the regime of isolated resonances. In 
section~\ref{S}, we study
Fano interference in the ``semiclassical regime'',
where many channels are strongly coupled to the interaction region. Even there,
we find a noticeable effect. Its strength is comparable to the weak 
localization correction in mesoscopic transport~\cite{MelBar99}. A summary is
given in section~\ref{C}.

In the present paper, the discussion is restricted to the case of orthogonal 
invariance, which allows to use the VWZ-integral~\cite{VWZ85} to compute 
autocorrelation functions.  For broken time-reversal symmetry, the 
corresponding result for the S-matrix element correlations has been given 
recently~\cite{SFS05}. It would allow to repeat the analysis of total cross 
section autocorrelation functions for that case.

\section{\label{F} The model}

We follow the approach adopted in~\cite{FyoAlh98,AFGIM03,GMI04} where resonant
and direct processes are treated separately by means of the Feshbach 
approach~\cite{Fes58,Fes62}: the Hilbert space of excited states is divided
into a subspace of bound states ($Q$-space), and its complement. The 
Schr\" odinger 
equation restricted to that complement is treated as an auxiliary scattering 
system. Choosing the bound space appropriately, it looses all resonant 
features. Its scattering solutions serve as a basis, in which the resonant 
scattering via the subspace of bound states is formulated. 

We consider the scattering problem (or the half collision problem) in 
a small energy interval, away from any thresholds, such that the number of
open channels, denoted by $M$, remains constant. Then, the partial decay 
amplitudes, defined in equation~(\ref{I:damp}), may be written as 
\begin{equation}
\fl t_a(E) = \la a|\alphaout\ra 
   + \la a|\, V^\dagger\; \frac{1}{E- H_\eff}\;
   \big( \, |\alphain\ra - \rmi\pi\; V\; |\alphaout\ra \, \big ) \qquad
H_\eff = H_0 -\rmi\pi\; V\, V^\dagger\; ,
\label{F:tamp}\end{equation}
where the vector $|\alpha\ra$ has been decomposed 
$|\alpha\ra = |\alphain\ra \oplus |\alphaout\ra$ according to the partitioning 
of the Hilbert space, above. Here, $H_0$ is a $N$$\times$$N$-matrix of the 
original Hamiltonian projected onto $Q$-space. The matrix $V$ 
contains the transition amplitudes between the basis states in $Q$-space,
and the scattering states $|c\ra$. It has $M$ column vectors (the 
channel vectors), each of dimension $N$. 

Note that $t_a(E)$ may be considered as a transition
amplitude between the state $|\alpha\ra$ and the scattering states $|a\ra$
of the auxiliary resonance-free scattering system. These are typically not
the free scattering solutions. To obtain the transition amplitudes into the 
free scattering states, one also needs the scattering matrix of the 
auxiliary scattering system.
Below, we focus on the total photo cross section, where such considerations 
are of minor importance.

To obtain the total photo cross section, one may sum over all partial cross
sections, or use a particular form of the optical theorem~\cite{Hel78a}. It
relies on the fact that the scattering solutions $|\Psi^{(c)}(E)\ra$ form a 
complete basis. Therefore:
\begin{eqnarray}
\fl \sigma(E) &=& 
   \sum_{c=1}^M \la\alpha|\Psi^{(c)}\ra\; \la\Psi^{(c)}(E)|\alpha\ra
 = \la\alpha|\, \delta(E-H)\, |\alpha\ra 
 = \frac{1}{\pi}\; {\rm Im}\, \la\alpha|\, G(E^-)\, |\alpha\ra \nonumber\\
\fl &=& \|\alphaout\|^2 - \pi^{-1}\; {\rm Im}\left[\,
   \big (\, |\alphain\ra + \rmi\pi\; V\; |\alphaout\ra \,\big )^\dagger\;  
   \frac{1}{E-H_\eff}\; 
   \big (\, |\alphain\ra - \rmi\pi\; V\; |\alphaout\ra \,\big )\, \right] \; .
\label{F:stot}\end{eqnarray}
Generally, we will assume that the column vectors of $V$ are pairwise 
orthogonal. If that is not the case, one may use a singular value 
decomposition~\cite{NumRec92}: $V= V_0\, s$, where $s$ is a unitary matrix.
It allows to replace $V$ by $V_0$, whose column vectors are pairwise 
orthogonal. Due to $V\, |\alphaout\ra = V_0\, |\alphaout_s\ra$ with 
$|\alphaout_s\ra= s\, |\alphaout\ra$, the expression~(\ref{F:stot}) remains 
unchanged except for a redefinition of the direct decay amplitudes.

\subsection{\label{FS} The statistical model}

For the derivation of the above expressions for $t_a(E)$ and $\sigma(E)$, 
we have used an arbitrary but fixed basis in $Q$-space. In that basis, we 
assume $V$, $\alphain$ and $\alphaout$ to be fixed, and take $H_0$ from the
Gaussian orthogonal ensemble (GOE). For the full collision problem, the 
statistical properties of the resulting S-matrix have been studied in 
detail~\cite{VWZ85,FyoSom97,GMW98}. Indirect photo decay has been 
studied in~\cite{FyoAlh98}, and the effects of direct decay paths has been 
considered in~\cite{AFGIM03,GMI04}.

For the description of decay processes, we have $3M+1$ independent parameters, 
which define the model:
\begin{itemize}
\item The norm of the column vectors of $V$ measure the coupling strengths to
the decay channels. 
\item The direct decay amplitudes $\alphaout_c= \la c|\alphaout\ra$ give 
the overlaps of the initial state with the scattering states of the
auxiliary scattering system. They describe direct decay processes, which may 
may lead to Fano interference.
\item The overlaps of $\alphain$ with the column vectors of $V$, {\it i.e.}
$\la c|\, V^\dagger\, |\alphain\ra$, may give rise to Fano interference, also. 
As we will see, this type of interference is not equivalent to the one before.
\item The norm of the component of $|\alphain\ra$, which is orthogonal to all
column vectors of $V$. That component is responsible for the indirect decay
processes.
\end{itemize}

\section{\label{M} The mapping formalism}

The mapping formalism has been introduced in~\cite{GMI04}. It allows to express
the partial decay amplitudes as well as the total photo cross section in terms
of an extended scattering matrix. This is convenient if analytical results for
the statistical scattering model are to be transfered to the present case.
Here, we will use it to compute the average (this section) and the 
autocorrelation function (section~\ref{S}) of the total photo cross section. 
In its original formulation, $|\alphain\ra$ is required to be orthogonal to 
the column vectors of $V$. Here, the formalism is generalized, to allow for 
non-zero overlaps between $|\alphain\ra$ and the channel vectors. This leads 
to the distinction between {\em external} and {\em internal} Fano interference.

Assume, the initial state $|\alphain\ra$ has overlap with the channel
vectors, and that channel vectors are coupled to the continuum with
non-vanishing transmission coefficients. In that case, we can always find a
$|\alpha^{\rm io}\ra$, such that
\begin{equation}
|\alphain\ra = |\alphain_0\ra +\pi\; V\; |\alpha^{\rm io}\ra \; ,
\label{M:alpio}\end{equation}
where $|\alphain_0\ra$ is orthogonal to all column vectors of $V$. Then, 
one can write for the transition amplitudes $t_a$ [equation~(\ref{F:tamp})]
and the total cross section [equation~(\ref{F:stot})]:
\begin{eqnarray}
\fl && t_a= \la a|\alphaout\ra + \la a|\, V^\dagger\; \frac{1}{E-H_{\rm eff}}\;
   \big ( \, |\alphain_0\ra -\rmi\pi\; V\; |\alphaout_+\ra \, \big ) \qquad
|\alphaout_\pm\ra = |\alphaout\ra \pm \rmi\; |\alpha^{\rm io}\ra \nonumber\\
\fl && \sigma(E)= \|\alphaout\|^2 - \pi^{-1}\; {\rm Im}\left[\,
  \big ( \, |\alphain_0\ra + \rmi\pi\; V\; |\alphaout_-\ra \, \big )^\dagger\;
  \frac{1}{E- H_\eff}\; 
  \big ( \, |\alphain_0\ra - \rmi\pi\; V\; |\alphaout_+\ra\, \big )\,
   \right] \; .
\end{eqnarray}
Following~\cite{GMI04}, this suggests to introduce the following extended
scattering matrix:
\begin{equation}
\fl S(E,\delta)= \One_{1+M} - 2\rmi\pi\; 
    W^\dagger\; \frac{1}{E- F_\eff}\; W \qquad
W= \big ( \, \alphain_0\; \delta/(2\pi)\, ,
         V \, \big ) \; ,
\label{M:Smat}\end{equation}
which contains only the orthogonal part of $|\alphain\ra$. Now, one has to
find transformation matrices $u(\delta)$ and $v(\delta)$, such that 
$S'(E,\delta)= u(\delta)\, S(E,\delta)\, v^T(\delta)$, may be used to construct 
the partial decay amplitudes, and the total photo cross section. One obtains 
the following answer:
\begin{equation}
\fl u(\delta) = \left( \begin{array}{c|c}
     \rmi & (\alphaout_-)^\dagger\; \delta/2 \\ \hline
              \begin{array}{c} 0      \\
                               \vdots \\
                               0 \end{array} &
              \begin{array}{ccc}   &        &   \\
                                   &  \One_{M} &   \\
                                   &        &  \end{array}
              \end{array} \right)\, \qquad
v(\delta) = \left( \begin{array}{c|c}
     \rmi & (\alphaout_+)^T\; \delta/2 \\ \hline
              \begin{array}{c} 0      \\
                               \vdots \\
                               0 \end{array} &
              \begin{array}{ccc}   &        &   \\
                                   & \One_M &   \\
                                   &        &  \end{array}
              \end{array} \right) \; .
\label{M:uv}\end{equation}
The transformation of $S(E,\delta)$ is reminiscent of an 
``Engelbrecht-Weidenm\" uller'' transformation \cite{EngWei73}, although for 
$\delta > 0$, $u(\delta)$ and $v(\delta)$ are not unitary. The partial decay 
amplitudes and the total photo cross section can be obtained from:
\begin{eqnarray}
\fl t_a(E) &=&  \frac{\la a|\alphaout_-\ra}{2} + 
   \lim_{\delta\to 0}\, \delta^{-1}\; S'_{a0}(E,\delta)
\label{M:tamp}\\
\fl \sigma(E) &=& \|\alphaout\|^2 -
   \frac{1}{2}\; {\rm Re}\, \la\alphaout_-|\alphaout_+\ra +
   \lim_{\delta\to 0}\, \frac{2}{\delta^2}\;
   {\rm Re}\big [\, 1 + S'_{00}(E,\delta) \,\big ] \; .
\label{M:stot}\end{eqnarray}

\paragraph{The average total cross section}
As a first application, we will compute the average total cross section
$\la\sigma(E)\ra$ in the center of the spectrum. To this end, we use the
fact that the average extended S-matrix is given by~\cite{VWZ85}:
\begin{equation}
\fl \la S_{ab}(0,\delta)\ra = \delta_{ab} \;
   \frac{1 - \lambda_a}{1+\lambda_a} \qquad
\lambda_a = \pi^2\; \rho_0\; \|\bm{v}_a\|^2\qquad
\lambda_0 = \frac{\delta^2}{4}\; \rho_0\; \|\alphain_0\|^2 \; ,
\end{equation}
where for $a=1,\ldots,M$, $\bm{v}_a$ are the respective column vectors of 
the coupling matrix $V$. The level density is denoted by $\rho_0$, {\it i.e.} 
the average level spacing is $\Delta = 1/(N\, \rho_0)$. As the average
S-matrix is diagonal, we obtain:
\begin{equation}
\la S'_{00}\ra = -\; \frac{1-\lambda_0}{1+\lambda_0} + \frac{\delta^2}{4}\;
   \la\alphaout_-|\, \bar S\, |\alphaout_+\ra \; ,
\end{equation}
where $\bar S$ is the average of the original scattering matrix (of dimension
$M$). From this it follows with equation~(\ref{M:stot}):
\begin{equation}
\la\sigma\ra = \|\alphaout\|^2 + \rho_0\; \|\alphain_0\|^2 + \frac{1}{2}\; 
   {\rm Re}\, \la\alphaout_-|\, (\bar S - 1)\, |\alphaout_+\ra \; .
\end{equation}
Provided, the components of $|\alphaout\ra$ and $|\alphain\ra$ are real, 
we may proceed a bit further. In equation~(\ref{M:alpio}), 
$|\alpha^{\rm io}\ra$ is chosen such that 
\begin{equation}
\la c|\alphain\ra = \pi\; \la \bm{v}_c|\alpha^{\rm io}\ra 
 = \pi \|\bm{v}_c\|\; \la c|\alpha^{\rm io}\ra \quad\Rightarrow\quad
\la c|\alpha^{\rm io}\ra = \sqrt{\rho_0/\lambda_c}\; \la c|\alphain\ra \; .
\end{equation}
Therefore:
\begin{eqnarray}
\la\sigma\ra &=& \|\alphaout\|^2 + \rho_0\; \|\alphain_0\|^2 - \sum_c
  \big (\, |\alphaout_c|^2 - \frac{\rho_0}{\lambda_c}\; |\alphain_c|^2\, \big )
   \frac{\lambda_c}{1+\lambda_c} \nonumber\\
 &=& \|\alphaout\|^2 + \rho_0\; \|\alphain\|^2 - \sum_c \big (\, 
   |\alphaout_c|^2 + \rho_0\; |\alphain_c|^2\, \big )\; 
   \frac{\lambda_c}{1+\lambda_c} \; .
\label{M:avstot}\end{eqnarray}
On the level of the average total cross section, it makes no difference, 
whether $\alphain$ has overlap with the external channel
region, or internal states (inside $Q$-space) which are connected to the
channel region via transmission coefficients. For later use, we define the
direct photo cross section as $\sigmadir = \|\alphaout\|^2$, and the indirect
photo cross section as $\sigmaind = \rho_0\; \|\alphain_0\|^2$.

\section{\label{L} Fano lineshapes: the limit of isolated resonances}

The effect of direct reaction paths is most evident in the limit of weak
coupling, where individual resonances can be observed. In this regime, the
resonances have asymmetric lineshapes (Beutler-Fano 
profiles)~\cite{Beu35,Fano61}, {\it i.e.} the cross section near resonance can
be parametrised as
\begin{equation}
\sigma(E) = \sigmadir\; \frac{|\eps +q|^2}{\eps^2+1} = \sigmadir\;
   \frac{(\eps +q_1)^2 + q_2^2}{\eps^2+1}\qquad
\eps = \frac{E-E_j}{\Gamma_j/2} \; .
\label{L:Fano}\end{equation}
While in the one channel case, the Fano parameter $q$ may be assumed real,
one needs a complex Fano parameter $q= q_1 +\rmi q_2$ in 
the case of many channels. The imaginary part of the Fano parameter lifts
the resonance, such that the cross section at its minimum is no longer zero.
As shown below, the background cross section (the cross section at large 
$|\eps|$) is $\sigmadir = \|\alphaout\|^2$, independent of the imaginary part 
of the Fano parameter.

In the limit of isolated resonances, the effective Hamiltonian $H_\eff$ can
be assumed diagonal in the eigenbasis of $H_0$. Using that basis in
equation~(\ref{F:stot}), we obtain:
\begin{equation}
\fl \sigma(E) = \|\alphaout\|^2 - \frac{1}{\pi}\; \sum_j\; {\rm Im}\left[
   \frac{(a_j+\rmi\, b_j)^*\; (a_j -\rmi b_j)} {E-E_j +\rmi\Gamma_j/2} 
   \right] \; .
\end{equation}
where $a_j = \la j|\alphain\ra$, $b_j= \pi\, \sum_c V_{jc}\; \alphaout_c$
and $\Gamma_j = 2\pi\sum_c V_{jc}^2$. If these parameters may be assumed 
real ({\it e.g.} due to time reversal invariance), we obtain in the vicinity 
of the $j$'th resonance:
\begin{equation}
\sigma(E) \approx \sigma_j(E) = \|\alphaout\|^2 - \frac{1}{\eps^2+1}\;
 \frac{2}{\pi\Gamma_j}\; {\rm Im}\big [\, (\eps-\rmi)\; (a_j-\rmi b_j)^2
 \, \big ] \; .
\end{equation}
This expression can be put into the form of equation~(\ref{L:Fano}) with
$\sigmadir = \|\alphaout\|^2$. This yields the following values for the 
real and the imaginary part of the Fano parameter:
\begin{equation}
q_1 = \frac{a_j\; b_j}{D_j}\qquad q_2= \sqrt{ 
   \left(1+ \frac{a_j^2}{D_j}\right)\; \left(1- \frac{b_j^2}{D_j}\right)} \; ,
\label{L:q}\end{equation}
where $D_j = \pi\; \|\alphaout\|^2\; \Gamma_j/2 \ge b_j^2$ due to the 
Schwarz inequality (it makes sure that $q_2$ is always real).

In the case of a regular cross section, it makes sense to study the individual
resonances and their lineshapes. However, if the lineshape ({\it i.e.} the
Fano parameter) fluctuates randomly from resonance to resonance, a statistical
analysis is more appropriate. Let us rewrite equation~(\ref{L:q})
in terms of $M+1$ independent normalized Gaussian random variables. We will 
take into account that $|\alphain\ra$ may have some overlap with channel 
vectors (the column vectors of $V$).
\begin{eqnarray}
\fl && V_{jc} \to s_c\; x_c \qquad 
a_j = \la j|\alphain_0\ra + \pi \sum_c V_{jc}\; \alpha^{\rm io}_c 
  \to s_0 \left(x_0 + \pi\; \sum_c x_c\; \frac{s_c\; \alpha^{\rm io}_c}{s_0} 
  \right) \nonumber\\
\fl && b_j = \pi\; \sum_c x_c\; \frac{s_c\; \alphaout_c}{s_0}\qquad
D_j = \pi^2\; \frac{\|\alphaout\|^2}{s_0^2}\; \sum_c x_c^2 \; s_c^2 \; .
\end{eqnarray}
Besides $\sigmadir = \|\alphaout\|^2$ and $\sigmaind \propto s_0^2$,
the following parameter combinations can be identified: $s_c$, 
$\, s_c\; \alphaout_c/s_0$, and $s_c\; \alpha^{\rm io}_c/s_0$, where 
$c=1,\ldots, M$. As these parameters come along with centered Gaussian random 
variables, their sign is irrelevant. We thus prefer to consider their squares:
\begin{eqnarray}
\fl && s_0^2= \frac{\|\alphain_0\|^2}{N}\qquad
s_c^2 = \la V_{jc}^2\ra \to \frac{1}{4\pi^2}\; \frac{T_c}{N\, \rho_0}
\nonumber\\
\fl && \frac{s_c^2\; |\alphaout_c|^2}{s_0^2} 
   \to \frac{|\alphaout_c|^2\; T_c}{4\,\pi^2\; \sigmaind}
   = \frac{\tau^{\rm out}_c}{4\pi^2}\qquad 
\frac{s_c^2\; |\alpha^{\rm io}_c|^2}{s_0^2} 
   \to \frac{|\alpha^{\rm io}_c|^2\; T_c}{4\,\pi^2\;\sigmaind}
   = \frac{\tau^{\rm io}_c}{4\pi^2}\; .
\label{L:taus}\end{eqnarray}
As we will see below, the autocorrelation function depends almost on the
same set of parameters. That means, an analysis of the autocorrelation 
function gives much the same information, with the advantage that its 
application is not restricted to the regime of isolated resonances.

\paragraph{One channel case} In that case
$b_j^2 = D_j$ so that the Fano parameter becomes real:
\begin{equation}
q_1 = \frac{a_j}{\pi\, \alphaout_1\, V_{j1}}
    = \frac{\la j|\alphain_0\ra}{\pi\;\alphaout_1\; V_{j1}} 
   + \frac{\alpha^{\rm io}_1}{\alphaout_1} \qquad
q_2 = 0 \; ,
\end{equation}
where we have again separated that part of $|\alphain\ra$ which is orthogonal 
to $V$ and that which is parallel. For precisely that case, the distribution 
of $q_1$ has been computed in~\cite{Ihra02}. The result is a shifted 
Lorentzian with width $g$ determined by:
\begin{equation}
g^2 = \frac{\lla\, \la j|\alphain_0\ra^2\, \rra}{\pi^2\, |\alphaout|^2\; 
   \la V_{j1}^2\ra}\; 
  = \frac{4\; \sigmaind}{|\alphaout|^2\; T_1} 
  = \frac{4}{\tau^{\rm out}_1}\; ,
\end{equation}
whereas the shift is given by
\begin{equation}
- \bar q = \frac{\alpha^{\rm io}_1}{\alphaout_1} 
  = \frac{\tau^{\rm io}_1}{\tau^{\rm out}_1} \; .
\end{equation}
That part of the work in~\cite{Ihra02}, which deals with time reversal 
invariant systems, is hence fully contained in the present model.

\section{\label{A} Autocorrelation function of the total photo cross section}

In section~\ref{L}, direct decay processes have been studied, as they affect
the lineshapes of isolated resonances. 
Once the resonances start to overlap, such an analysis 
is no longer possible. Here, we study the interference effects on the 
autocorrelation function. This quantity has the advantage, that there are no
restrictions on the coupling strengths to decay channels (transmission 
coefficients). This allows to study Fano interference even in the regime of
strongly overlapping resonances (section~\ref{S}).

\paragraph{Internal versus external Fano interference}
Here, we will study the autocorrelation function of the total photo cross
section~(\ref{F:stot}), with the help of the mapping formalism and the 
VWZ-integral~\cite{VWZ85}. The latter provides an analytical expression for
the correlation function between two matrix elements of the (extended)
scattering matrix in equation~(\ref{M:Smat}). With this, it is possible to
compute correlation functions between partial decay amplitudes, as well as
between total cross sections. It is also possible to choose different initial
states $\alpha$, if that would be of interest. In the case of 
two total cross sections, we would consider the correlation function
\begin{equation}
\fl \qquad C[\sigma_1,\sigma_2](w) = 
   \lla\sigma_1(E-w\, \Delta/2)\, \sigma_2(E+w\, \Delta/2)\rra -
   \lla\sigma_1(E)\rra \; \lla\sigma_2(E)\rra \; ,
\end{equation}
where $\la\ldots\ra$ denote a spectral ane/or ensemble average, as 
appropriate, and $\Delta$ the mean level (resonance) spacing. Here, 
$\sigma^{(1)}$ ($\sigma^{(2)}$) denotes the total photo
cross section for the initial state $|\alpha_1\ra$ ($|\alpha_2\ra$). Usually,
we find it more convenient to consider the correlation function in the time
domain:
\begin{equation}
\hat C[\sigma_1,\sigma_2](t) = \int\rmd w\; \rme^{2\pi\rmi\, wt}\;
   C[\sigma_1,\sigma_2](w) \; .
\end{equation}
Due to the fact that $C[\sigma_1,\sigma_2](w)$ is defined on the unfolded
energy axis, $t$ measures time in units of the Heisenberg time.

For simplicity, and because this case is typically considered in the 
literature, the following discussion will be restricted to the
autocorrelation function of the total photo cross section.
Using for $\sigma$ the expression~(\ref{M:stot}) from the mapping formalism,
(section~\ref{M}) one obtains:
\begin{equation}
C[\sigma] = \lim_{\delta\to 0} \frac{1}{\delta^4}\;
   C[S_{00}'+S_{00}^{\prime\ast},S_{00}'+S_{00}^{\prime\ast}]
 = 2\; \lim_{\delta\to 0} \frac{1}{\delta^4}\;
      {\rm Re}\, C[S_{00}',S_{00}^{\prime\ast}] \; ,
\end{equation}
and for its Fourier transform (see reference~\cite{GS02}):
\begin{equation}
\hat C[\sigma] = \lim_{\delta\to 0} \frac{1}{\delta^4}\;
      \hat C[S_{00}',S_{00}^{\prime\ast}] \; .
\end{equation}
The auxiliary scattering matrix $S'(E,\delta)$ is defined within the mapping
formalism [see equations~(\ref{M:Smat}) and~(\ref{M:uv})]. For its matrix 
element $S_{00}'$, one obtains:
\begin{equation}
\fl S_{00}' = - S_{00} + \frac{\rmi}{2}\, \delta\; \sum_{a=1}^M \big [\,
   (\alphaout_-)_a^*\; S_{a0} + (\alphaout_+)_a\; S_{0a}\, \big ]
   + \frac{\delta^2}{4}\; \sum_{a,b=1}^M (\alphaout_-)_a^*\; (\alphaout_+)_b\;
   S_{ab} \; .
\end{equation}
This gives $\hat C[\sigma](t)$ in terms of a linear combination of
correlation functions between S-matrix elements, where the 
VWZ-integral~\cite{VWZ85} applies. Due to equation~(\ref{aV:VWZ}) many such 
correlation functions vanish. After a bit of algebra, one arrives at the
following result:
\begin{eqnarray}
\fl &&\hat C[\sigma] = \sigmaind^2\; \mathcal{I}\left\{
 \left|\textstyle
    2\,\Delta_0 - \frac{1}{2}\; \sum_{a=1}^M \tau_a\; \Delta_a \right|^2
 + 4\, \Pi_{00}
 +\, \frac{1}{2}\; \sum_{a=1}^M \big [\,  \tau_a +\tau_a^* + \tau_a^- +
      \tau_a^+\, \big ]\; \Pi_{0a} \right. \nonumber\\
\fl &&\qquad\left. +\, \frac{1}{8}\; \sum_{a,b=1}^M
   \big [\, \tau_a^-\; \tau_b^+ + \tau_a\; \tau_b\, \big ]\; \; \Pi_{ab}
\right\} 
\label{A:Cphgen}\\
\fl &&\tau_a = (\alphaout_-)_a\; (\alphaout_+)_a^*\; T_a/\sigmaind \qquad
\tau_a^\pm = |\, (\alphaout_\pm)_a\, |^2\; T_a/\sigmaind \; .
\nonumber\end{eqnarray}
If the Fano interference is purely external, then
$\tau_a= \tau_a^\pm\, :\, $ real, and we obtain the result from~\cite{GMI04}:
\begin{equation}
\fl \hat C[\sigma](t) = {\cal I} \left\{ \left(\textstyle
  2\, \Delta_0 - \frac{1}{2}
  \sum_{a=1}^M \tau_a\, \Delta_a\right)^2 + 4\; \Pi_{00} + 2\sum_{a=1}^M
  \tau_a\, \Pi_{0a} + \frac{1}{4}\sum_{a,b=1}^M \tau_a\tau_b\, \Pi_{ab}
  \right\} \; ,
\label{A:Cph}\end{equation}
where $\tau_c= |\alphaout_c|^2\, T_c /\sigmaind = \tau^{\rm out}_c$ as given
in equation~(\ref{L:taus}). The parameters $\tau_c$ and
$\tau_c^\pm$ are the statistical variants of the Fano parameter; see 
section~\ref{L}. Eventually, we call them
``statistical Fano parameters'' for short. If Fano interference is purely 
internal, then
$\tau_a^\pm = |\alpha^{\rm io}_a|^2\, T_a/\sigmaind = \tau^{\rm io}_a$, 
$\tau_a = - \tau_a^\pm$.
Therefore:
\begin{equation}
\hat C[\sigma](t) = {\cal I} \left\{ \left(\textstyle
  2\, \Delta_0 + \frac{1}{2}
  \sum_{a=1}^M \tau^\pm_a\, \Delta_a\right)^2 + 4\; \Pi_{00}
  + \frac{1}{4}\sum_{a,b=1}^M \tau^\pm_a\tau^\pm_b\, \Pi_{ab}
  \right\} \; .
\end{equation}

\section{\label{S} Limit of many open channels}

We consider two cases. The first is the {\em absorptive limit}. There,
all transmission coefficients go to zero, while the number of channels goes
to infinity: $M\to\infty$. The limits are taken in such a way that the sum 
over all transmission coefficients remains finite: $\sum_c T_c = T_{\rm sum}$. 
In that case, the resonance widths stop fluctuating~\cite{EriMay66}. In the 
second case, we assume that all transmission coefficients are equal to one, 
while the number of channels is large but finite. We refer to that case as the 
{\em semiclassical regime}.

\paragraph{Absorptive limit}
In this case, the autocorrelation function can be computed in closed form (all
integrals can be solved, see~\ref{aV}), and one obtains:
\begin{equation}
\hat C[\sigma] = \sigmaind^2\; \rme^{-T_s\, t}\left\{ 
   \big (\, 1 - \frac{1}{4}\; \sum_a \tau_a\, \big )^2\; 
      \big [\, 1- b_2(t)\, \big ] + 2 + C_0 \right\} \; ,
\end{equation}
where the constant $C_0$ depends on the statistical Fano parameters. Hence,
in the absence of internal Fano interference, one can obtain a perfect 
exponential autocorrelation function by properly tuning $\sum_a \tau_a \to 4$.

\paragraph{Semiclassical regime}
For simplicity, we focus on the two extreme cases. In the first case, the 
decay is dominated by indirect processes, such that the autocorrelation function
becomes:
\begin{equation}
\hat C[\sigma](t) = 4\; \sigmaind^2\; {\cal I}^{(M)}\; (\Delta_0^2 + \Pi_{00})
\; ,
\end{equation}
while in the second case, it is dominated by direct processes. This results
in 
\begin{equation}
\hat C[\sigma](t) = \sigmaind^2\;
   \frac{\sum_{a,b} (\tau_a^-\, \tau_b^+ + \tau_a\, \tau_b)}{8}\;
   {\cal I}^{(M)}\; \Pi_{11} \; .
\end{equation}
As can be seen from equation~(\ref{aV:VWZ}), the autocorrelation function is
then proportional to the autocorrelation function of a diagonal S-matrix
element (note that all transmission coefficients are equal to one). Thus, very
large statistical Fano parameters ({\it i.e.} the dominance of the direct 
processes) result in a photo cross section which has the same statistical
properties as a full scattering cross section 
$\sigma_{\rm tot}(E)$~\cite{GS02}. Below, we will thus compare 
$\hat C[\sigma](t)$ for indirect decay with $\hat C[\sigma_{\rm tot}](t)$ for 
direct decay. Note that for large $M$, one expects the autocorrelation 
function in both cases to be dominated by an exponential decay with 
$\rme^{-Mt}$~\cite{EngWei73,BluSmi88,Aga99}.

\begin{figure}
\input{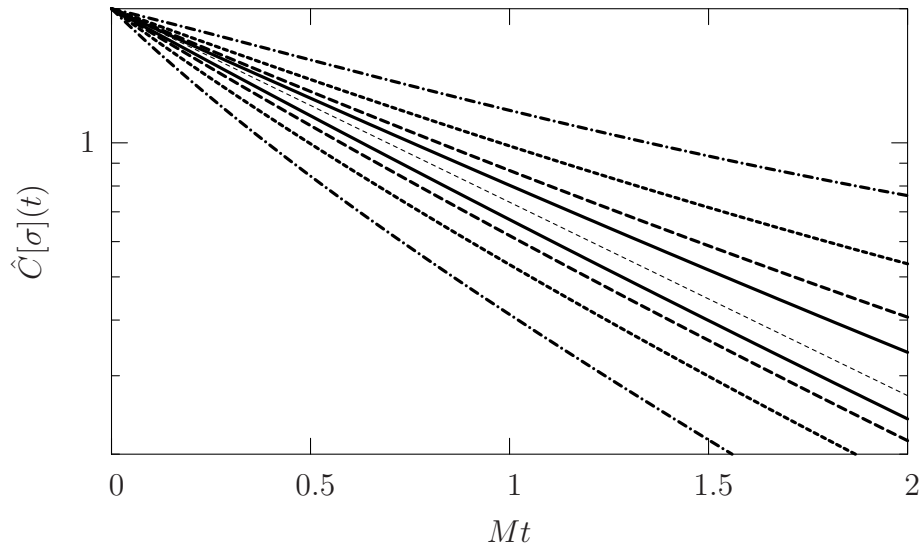}
\caption{The autocorrelation function $\hat C[\sigma](t)$ versus $Mt$, in the 
absence of direct processes (indirect cross section) compared to the same 
quantity where the direct processes dominate. The curves are computed 
numerically, on the basis of the VWZ integral, see~\ref{aV}.
All transmission coefficients are taken equal to one, and the number of 
channels is varied between $M=16$ (solid line), 8 (dashed line), 4 (dotted 
line), and 2 (dash-dotted line). The thin dotted line shows the exponential 
decay: $2\, \exp(- Mt)$.}
\label{S:f1}\end{figure}

In figure~\ref{S:f1} we simply compare the resulting autocorrelation 
functions, varying the number of channels. A semilog plot is used, and on the
ordinate $Mt$ is given, such that differences to the exponential decay are 
easier recognized. The autocorrelation functions are normalized, such that 
$\hat C[\sigma](0) = \hat C[\sigma_{\rm tot}](0) = 2$. In the case of indirect 
decay, the autocorrelation function lies above the purely exponential decay 
(thin dotted line), approaching the exponential as $M$ increases. In the case 
of direct decay, the autocorrelation function approaches the exponential from 
below.

\begin{figure}
\input{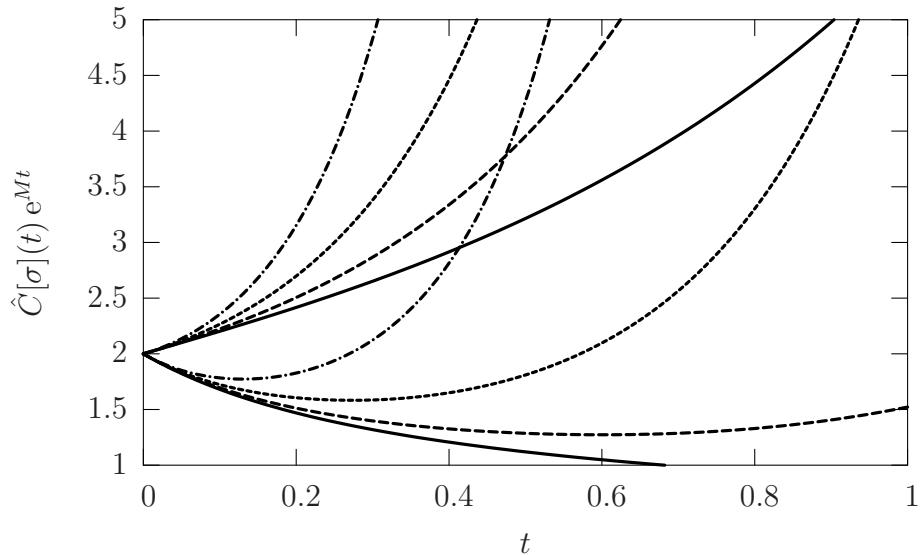}
\caption{The same autocorrelation functions as in figure~\ref{S:f1}, but 
divided by $\rme^{-Mt}$, and plotted versus $t$. Again, the number of channels
is varied between $M=2$ (dash-dotted lines), $M=4$ (dotted lines), $M=8$ 
(dashed lines), and $M=16$ (solid lines). The autocorrelation functions for 
indirect decay have a positive slope at $t=0$, while in the opposite case,
they have a negative slope.}
\label{S:f2}\end{figure}

In figure~\ref{S:f2} the same autocorrelation functions are shown, again.
However, here, we divide by the classical expectation $\exp(-Mt)$ and we plot
the autocorrelation functions versus time $t$ (in units of the Heisenberg 
time). In this way, differences to the classical expectation are strongly
enhanced. For indirect decay, the autocorrelation function has
been computed semi-classically in~\cite{Aga99}, with the result: 
$\hat C[\sigma](t)= 2\, (1+t)\; \rme^{-Mt}$. For the autocorrelation function 
of total scattering cross sections, a similar result is not available
(to the best of the author's knowledge). One may however find a 
connection to the weak localization correction of mesoscopic 
transport~\cite{MelBar99}, because the integral over the autocorrelation 
function in the time domain, gives the average value of the modulus squared 
of the corresponding S-matrix element~\cite{GS02}:
\begin{equation}
\int_0^\infty\rmd t\; \hat C[\sigma](t) = \la |S_{aa}|^2\ra = \frac{2}{M+1}\; .
\label{S:wlc}\end{equation}
\mbox{}\\

\begin{table}
\begin{tabular}{|r|c|c|c|}
\hline
M & a & b & c \\
\hline\hline
 2 & 1.99934(54) & 0.076(48) & 0.1(15) \\
\hline
 4 & 2.00018(35) & 1.991(30) & 3.67(87) \\
\hline
 8 & 2.00029(44) & 5.987(37) & 5.0(10) \\
\hline
16 & 2.00057(21) & 13.953(17) & 8.74(49) \\
\hline
\end{tabular}
\caption{The coefficients $a,b,c$ in equation~(\ref{S:coeffs}) as obtained from
a polynomial fit to the $\hat C[\sigma](t)$ as plotted in figure~\ref{S:f2}. 
The digits in brackets give the estimated error from the 
$\chi^2$-fit~\cite{NumRec92}. The first column gives the number of channels.}
\label{S:t1}\end{table}

To study the behavior of the autocorrelation functions $\hat C[\sigma](t)$
(indirect decay) and $\hat C[\sigma_{\rm tot}](t)$ more quantitatively, we
perform a polynomial fit to the numerical results, as shown in 
figure~\ref{S:f2}. To this end we use the following general expression:
\begin{equation}
\hat C[\sigma](t) \sim (2 + a\, t + b\, t^2 + c\, t^3 + \ldots )\; \rme^{-Mt}
\; .
\label{S:coeffs}\end{equation}
The results of the fits are given in the tables~\ref{S:t1} and~\ref{S:t2}. We 
used
a polynomial of fifth order for the fit, but the values for the higher order
coefficients had too large errors to be of any use. On the basis
of the values found, one may conjecture:
\begin{eqnarray}
\hat C[\sigma](t) &\sim & (2 + 2 t + (M-2) t^2)\; \rme^{-M\, t} \\
\hat C[\sigma_{\rm tot}](t)    &\sim & (2 - 4 t + (M+4) t^2)\; \rme^{-M\, t}
\; .
\label{S:sigtot}\end{eqnarray}
In the case of indirect decay, we obtain agreement with~\cite{Aga99} up to
the linear term. Higher order semiclassical corrections to the indirect photo
cross sections have not been considered, so far. In the second case, we may
actually integrate equation~(\ref{S:sigtot}) to recover the weak localization
correction:
\begin{equation}
\int_0^\infty\rmd t\; \hat C[\sigma_{\rm tot}](t) \sim \frac{2}{M} 
   - \frac{4}{M^2} + \frac{2}{M^2} + \ldots
\end{equation}
in agreement with an expansion of equation~(\ref{S:wlc}) in inverse powers of 
$M$. This may of course just be a coincidence. For a definite answer, one 
should try to integrate ${\cal I}^{(M)}\, \Pi_{11}$ in closed form.

\begin{table}
\begin{tabular}{|r|c|c|c|}
\hline
M & a & b & c \\
\hline\hline
 2 & -4.00109(70) & 10.101(62) & -27.1(19) \\
\hline
 4 & -4.00039(48) & 12.028(41) & -32.5(12) \\
\hline
 8 & -3.99930(30) & 15.945(25) & -46.21(72) \\
\hline
16 & -3.99951(21) & 23.948(18) & -77.58(49) \\
\hline
\end{tabular}
\caption{The coefficients $a,b,c$ in equation~(\ref{S:coeffs}) as obtained from
a polynomial fit to $\hat C[\sigma_{\rm tot}](t)$ as plotted in 
figure~\ref{S:f2}.  The digits in brackets give the estimated error from the 
$\chi^2$-fit~\cite{NumRec92}. The first column gives the number of channels.}
\label{S:t2}\end{table}

\section{\label{C} Conclusions}

This work has started out from a recent formulation of a statistical model
for half collision processes~\cite{AFGIM03,GMI04}. The principal aim was to
investigate the possibility to detect and quantify Fano interference in cases
where the total photo cross section shows irregular behavior (that may be
randomly fluctuating lineshapes and/or overlapping resonances). The analysis
of the Fano-parameter distribution~\cite{CWB01,Ihra02} may be appropriate, as
long as the resonances are well isolated. However, 
the cross section autocorrelation function, can always be analysed.
Surprisingly, it provides practically 
the same information (statistical Fano parameters) as the analysis of the 
Fano-parameter distribution. This allowed us to investigate Fano interference
in the regime of strongly overlapping resonances.
There, Fano interference effects are much weaker (on the
level of the weak localization correction).

One might have doubts, whether Fano interference can also be detected in
practice, when many channels are strongly coupled to the interaction region.
Given a dynamical system, the main question is, how much averaging over 
different initial conditions and/or different samples is necessary in order 
to obtain the desired information.
To clarify that point, numerical simulations with the open kicked 
rotor~\cite{FyoSom00,Oss03,TwoSch03} are currently under investigation.

Fano interference has turned into a versatile tool in mesoscopic physics,
where it is used to measure dephasing and decoherence 
times~\cite{CWB01,Kob04}. So far, only single Fano resonances in single mode
systems have been applied. It would be desirable, to be able to measure 
dephasing and decoherence times also in multi-mode systems, from irregular
cross sections. As a first step in that direction, one should include 
dephasing and decoherence into the statistical model.

\ack
Very fruitful discussions with D. F. Martinez, H. Schomerus and B. Mehlig
are greatfully acknowledged.

\begin{appendix}

\section{\label{aV} The Verbaarschot-Weidenm\" uller-Zirnbauer integral}

In the case that $H_0$ is taken from the GOE, the correlation function between
two scattering matrix elements from equation~(\ref{M:Smat}) is given by the
VWZ-integral~\cite{VWZ85}.
Fourier-transformed into the time domain, it reads~\cite{GS02}:
\begin{equation}
\fl \qquad
\hat C[S_{ab},S_{cd}^*](t) = {\cal I} \left\{ 4\delta_{ab}\delta_{cd}\, T_a
  T_c\; \Delta_a\, \Delta_c +
  2(\delta_{ac}\delta_{bd} + \delta_{ad}\delta_{bc})\; T_a\, T_b\; \Pi_{ab}
  \right\} \; .
\label{aV:VWZ}\end{equation}
Here, the $T_a$ are the transmission coefficients, and we used the following
abbreviations:
\begin{eqnarray}
\fl {\cal I} &=& \int_{\max(0,t-1)}^t\rmd r\int_0^r\rmd u\;
  \frac{(t-r)(r+1-t)}{(2u+1)(t^2-r^2+x)^2}\; \prod_{e=1}^M
  \frac{1-T_e (t-r)}{\sqrt{1+2T_e r + T_e^2 x}} 
\label{aV:I}\\
\fl \Delta_a &=& \sqrt{1-T_a} \left( \frac{r+T_a x}{1+2T_a\, r +T_a^2\, x} +
  \frac{t-r}{1- T_a(t-r)}\right)\qquad
x= u^2\, \frac{2r+1}{2u+1}\nonumber\\
\fl \Pi_{ab} &=& 
  \frac{T_a T_b\, x^2 + [T_a T_b r+ (T_a+T_b) (r+1) -1] x +(2r+1) r}
  {(1+2T_a\, r + T_a^2\, x)(1+2T_b\, r +T_b^2\, x)}
  + \frac{(t-r)(r+1-t)}{[1-T_a(t-r)]\, [1-T_b(t-r)]} \; .
\nonumber\end{eqnarray}

\paragraph{Absorptive limit}
There, all transmission coefficients go to zero, while the number of channels 
goes to infinity: $M\to\infty$. Both limits are taken in such a way that 
$\sum_c T_C = T_{\rm sum}$ remains finite. In that case, one find:
\begin{eqnarray}
\Delta_a &\to& \Delta_0 = t \nonumber\\
\Pi_{ab} &\to& \Pi_{00} = (t-r) (r+1-t) - x + (2r+1)\, r 
\label{aV:Erilims}\\
{\cal I} &\to& \rme^{-T_s\, t}\; {\cal I}^{(0)} \qquad
{\cal I}^{(0)} = \int_{\max(0,t-1)}^t\rmd r\int_0^r\rmd u\;
  \frac{(t-r)(r+1-t)}{(2u+1)(t^2-r^2+x)^2}\; .
\nonumber\end{eqnarray}
The integrals ${\cal I}^{(0)}$ and ${\cal I}^{(0)}\, \Pi_{00}$ can be 
calculated in closed form~\cite{StoSch04}. This gives:
\begin{equation}
{\cal I}^{(0)} = \frac{1- b_2(t)}{4\, t^2}\;  \qquad
{\cal I}^{(0)}\; \Pi_{00} = \frac{1}{2} \; . 
\end{equation}
For the GOE spectrum, the two point form factor $b_2(t)$ is given by
\begin{equation}
1-b_2(t) = 2 t -t\; \ln(2 t+1) + \theta(t-1)\; \big [\, 
   2 - 2t + t\; \ln(2t-1)\, \big ] \; .
\end{equation}
This gives for the correlation function:
\begin{equation}
\fl \hat C[S_{ab},S_{cd}^*](t) = \rme^{-T_s\, t}\; \big [\, 
   \delta_{ab}\delta_{cd}\, T_a T_c\; (1-b_2(t)\, ) +
  (\delta_{ac}\delta_{bd} + \delta_{ad}\delta_{bc})\; T_a\, T_b\, \big ] \; .
\end{equation}

\paragraph{The semiclassical regime}
There, all transmission coefficients are set equal to one 
$\forall c : T_c = 1$. Typically the number of channels $M$ is assumed to be
large but finite. However, the following relations really hold for arbitrary
$M$. In that case, one finds:
\begin{equation}
\Delta_a \to \left\{ \begin{array}{cc} \Delta_0 = t &: T_a\to 0\\
   \Delta_1 = 0 &: T_a\to 1 \end{array} \right. \qquad
\Pi_{ab} \to \left\{ \begin{array}{cl} \Pi_{00} &: T_a,T_b \to 0\\
   \Pi_{01} &: T_a\to 0, T_b\to 1 \\
   \Pi_{11} &: T_a,T_b \to 1
\end{array} \right. \; ,
\end{equation}
where $\Pi_{00}$ is given in equation~(\ref{aV:Erilims}), and
\begin{equation}
\Pi_{01} = t\qquad 
\Pi_{11} = \frac{x+r}{1+2r+x} + \frac{t-r}{1+r-t} \; .
\end{equation}
Finally,
\begin{equation}
\fl {\cal I} \to {\cal I}^{(M)} = \int_{\max(0,t-1)}^t\rmd r\int_0^r\rmd u\;
  \frac{(t-r)(r+1-t)}{(2u+1)(t^2-r^2+x)^2}\; \left(
  \frac{1+r-t}{\sqrt{1+2 r + x}}\right)^M \; . 
\end{equation}
The resulting integrals are computed numerically.

\end{appendix}

\section*{References}

\bibliographystyle{unsrt}

\bibliography{/home/gorin/Bib/deco,/home/gorin/Bib/amol,/home/gorin/Bib/qdot,/home/gorin/Bib/rom,/home/gorin/Bib/semic,/home/gorin/Bib/stas,/home/gorin/Bib/books}


\end{document}